\documentclass[twocolumn,showpacs,preprintnumbers,amsmath,amssymb,prb,floatfix]{revtex4}

\usepackage{graphicx}
\usepackage{dcolumn}
\usepackage{bm}
\usepackage{array}
\usepackage{amsmath,amssymb}

\newcommand{\be}{\begin{eqnarray}}
\newcommand{\ee}{\end{eqnarray}}

\begin{document}
\title{Zero modes and  the edge states of the honeycomb lattice  }
\author{ Mahito Kohmoto$^{1}$ and  Yasumasa Hasegawa$^2$}
\affiliation{
$^1$Institute for Solid State Physics, University of Tokyo, 
5-1-5 Kashiwanoha, Kashiwa, Chiba 277-8581, Japan\\
$^2$Department of Material Science,
Graduate School of Material Science,
University of Hyogo,
 Ako, Hyogo 678-1297, Japan
}
\date{\today}
\begin{abstract}
The honeycomb lattice in the cylinder geometry with zigzag edges, bearded edges, 
zigzag and bearded edges (zigzag-bearded), and armchair edges are studied.
The tight-binding model with nearest-neighbor hoppings is used.
Edge states  are obtained analytically for  these edges except the armchair edges. 
It is shown, however, that  edge states for the armchair edges exist when the the system is anisotropic. 
These states have not been known previously.
We also find strictly localized states, uniformly extended states 
and states with macroscopic degeneracy.
\end{abstract}


\maketitle

\section{Introduction}
Monolayer graphite, called graphene, was fabricated recently\cite{novo,zhang,berger2006}
 and novel
physical properties have been expected to be seen. In fact, integer quantum Hall
effect   has been reported \cite{novo,zhang}. 

\begin{figure}[t]
\begin{center}
\includegraphics[width=70mm]{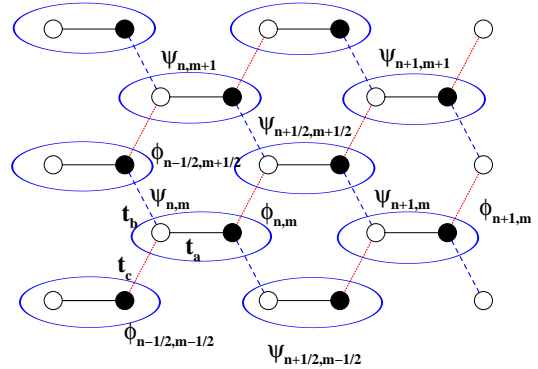}
\caption {The honeycomb lattice. Open and closed circles shows sublattice A and B, respectively.
 $t_a$, $t_b$, and $t_c$ are hopping integrals. }
\label{honey00}
\end{center}
\end{figure}

\begin{figure}[p]
\begin{center}
\begin{tabular}{c}
\includegraphics[width=55mm]{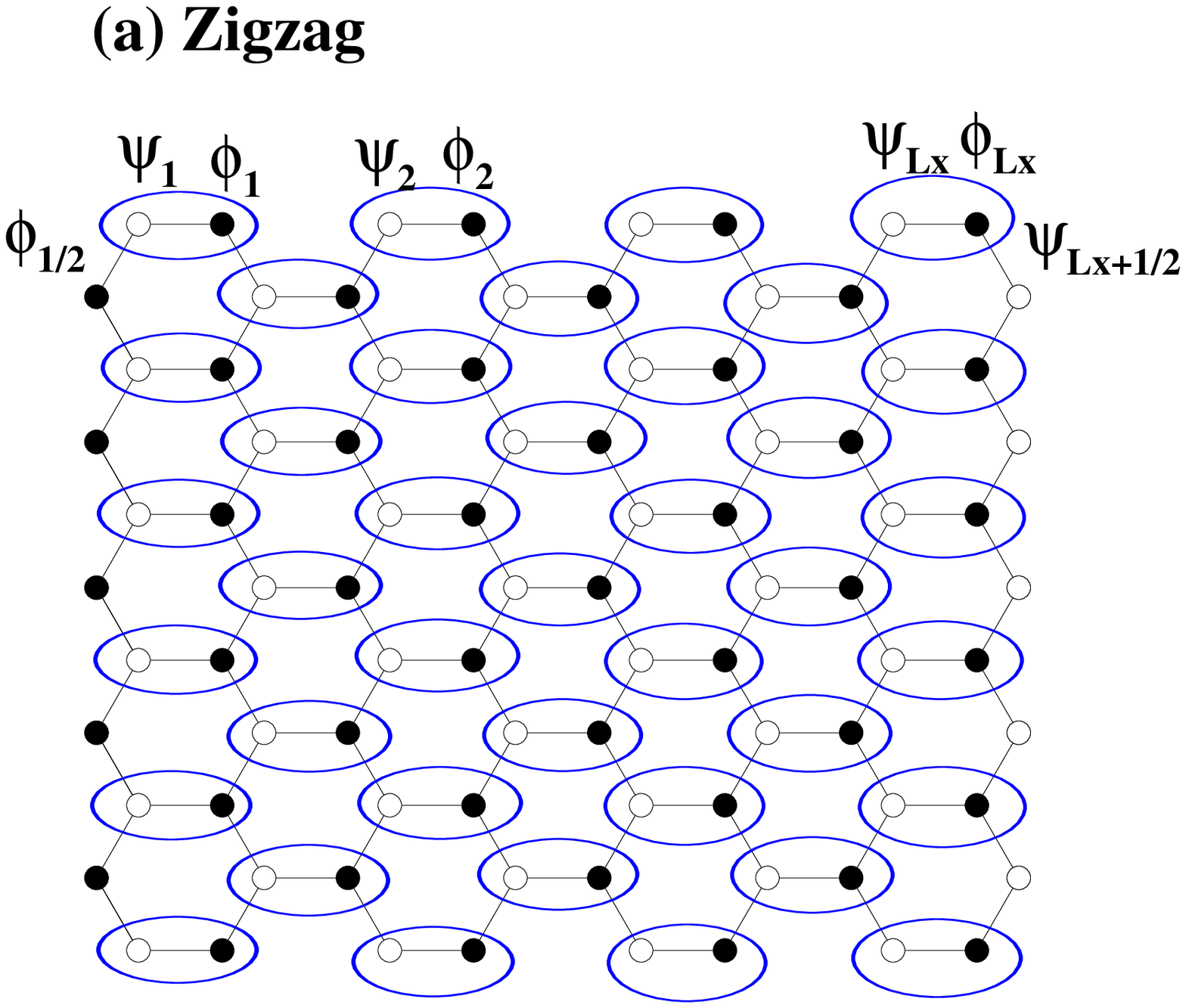}
\\
\\
\includegraphics[width=55mm]{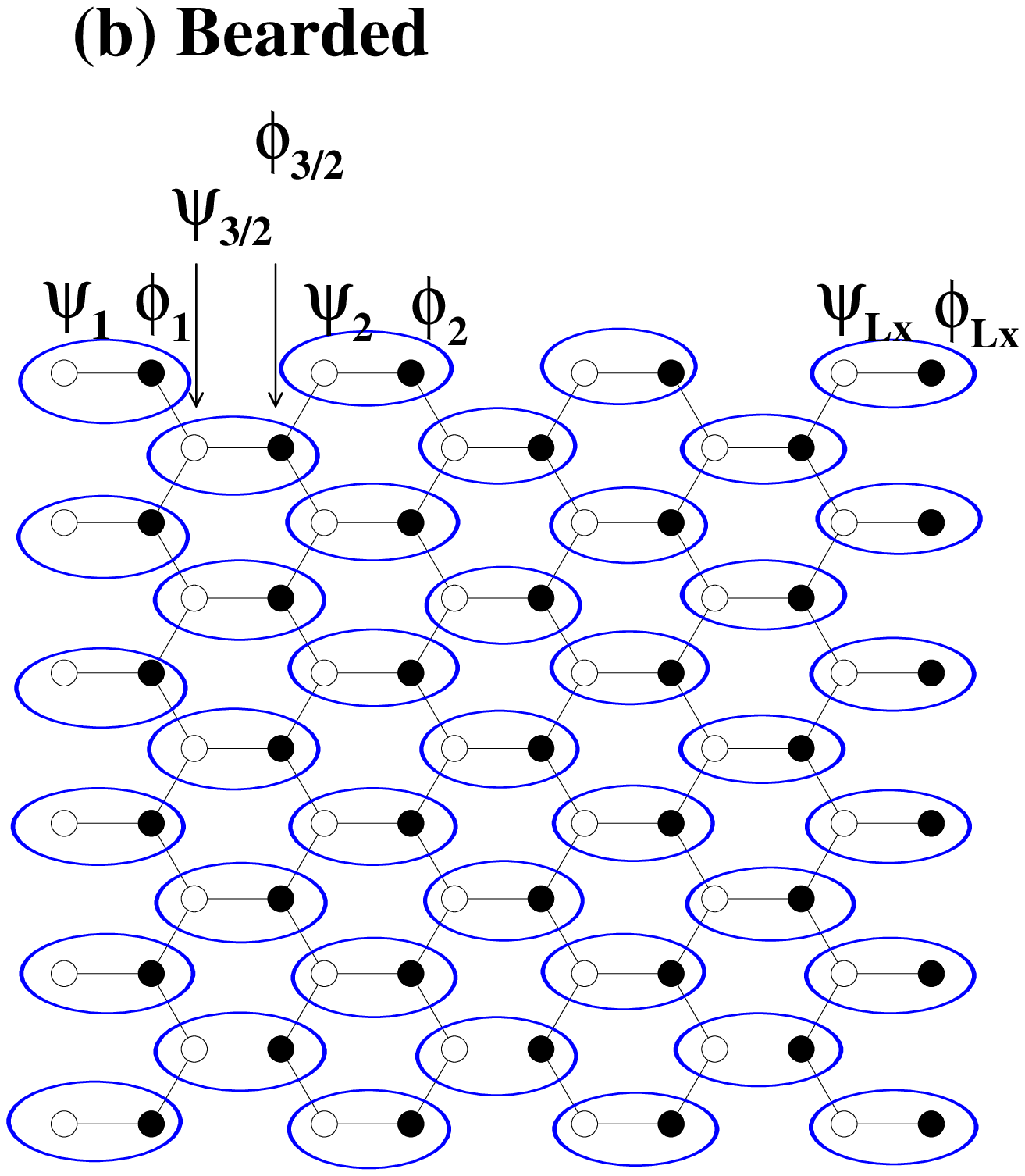}
\\
\includegraphics[width=55mm]{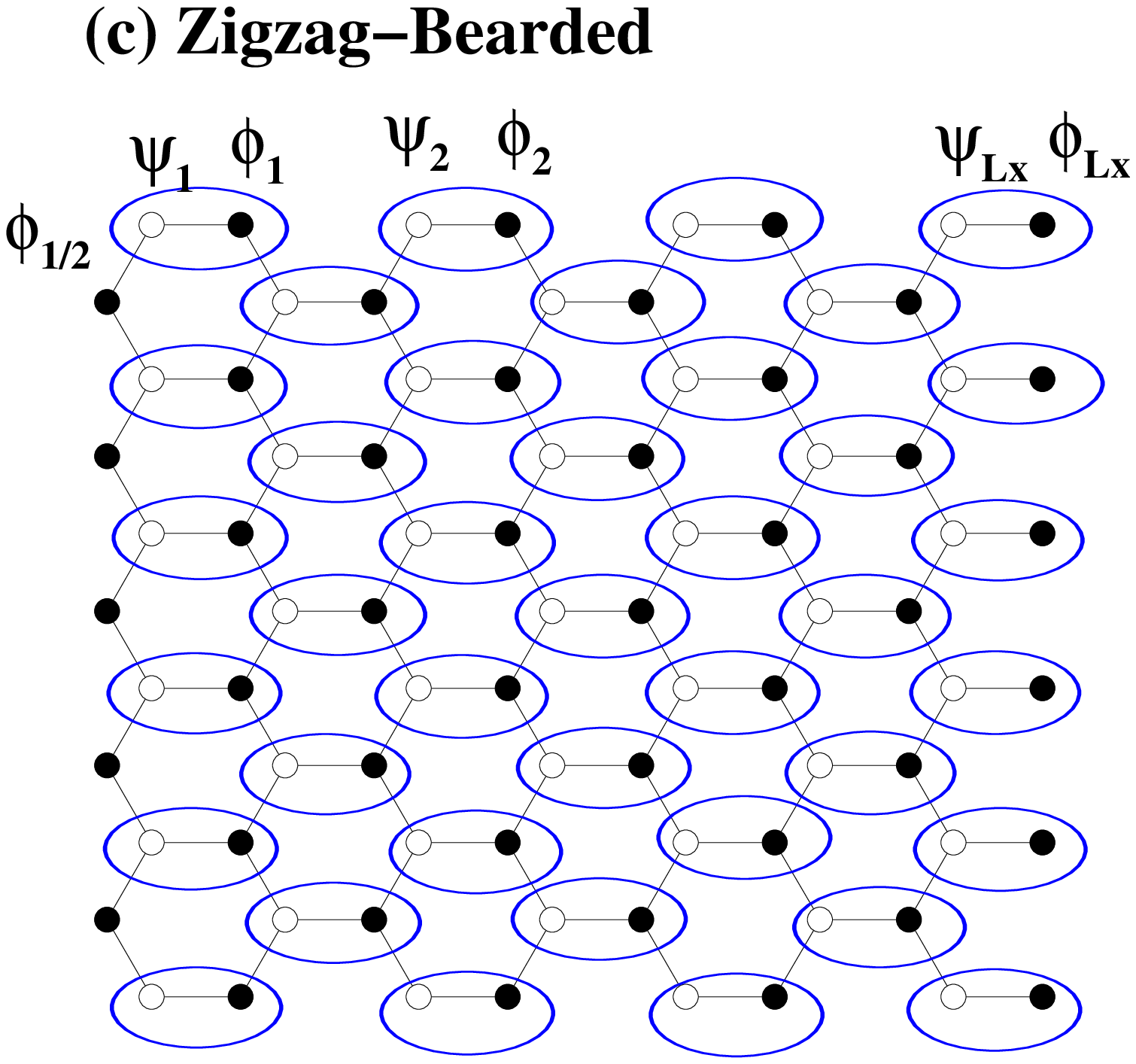}
\\
\includegraphics[width=55mm]{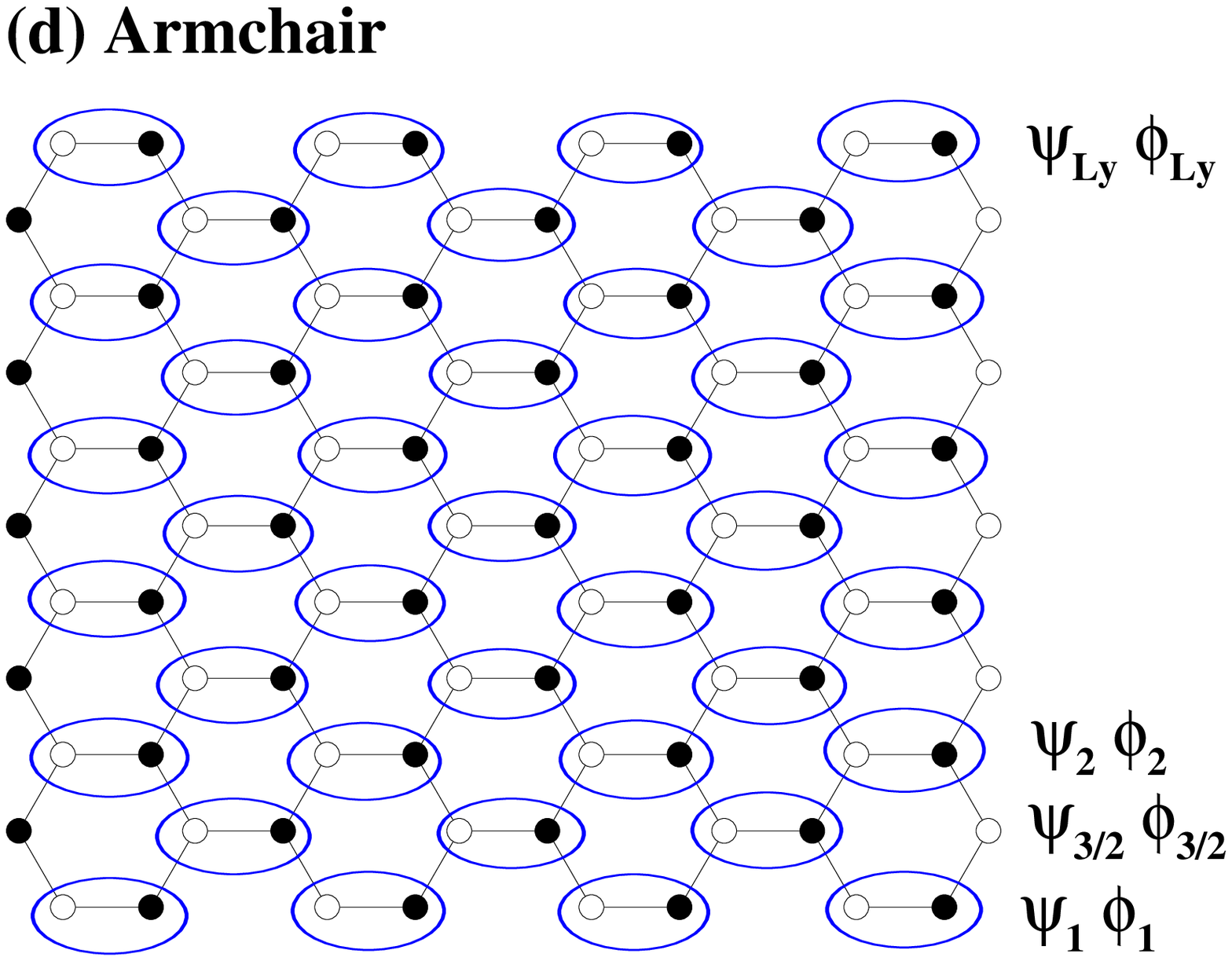}
\end{tabular}
\caption {The honeycomb lattices with (a)  zigzag ($L_x=4$), (b) bearded ($L_x=4$), 
(c) zigzag-bearded ($L_x=4$), and (d) armchair ($L_y=6$) edges, which we call Zigzag,
 Bearded, Zigzag-Bearded, and Armchair, respectively. }
\label{fig2}
\end{center}
\end{figure}

In this paper we report  a systematic study of the zero modes and the corresponding edge 
states of the honeycomb lattice which is shown in Fig. \ref{honey00}. 
The cylindrical geometry 
is taken and  thus two edges are present. 
We consider three types of edges:  zigzag, bearded, and armchair which are shown in Fig.~\ref{fig2}.
Two edges of the same type  can form a cylinder. In
addition, zigzag and bearded edges can form a cylinder.  We call these
as Zigzag, Bearded, Armchair, and Zigzag-Bearded, respectively.
An armchair edge and a zigzag edge(or a bearded edge) can not form a pair of edges for a cylinder. 
When only nearest-neighbor hoppings are taken, the zero-energy edge states  for 
Zigzag, Bearded, and Zigzag-Bearded are obtained. They are localized near 
the edges with the localization length 
\be
\xi = \frac{1}{2 |\log t |} ,
\label{localization}
\ee
 where 
 \be
t=\sqrt{2(1+\cos k_y)}=2 |\cos \frac{k_y}{2} |,
\label{intt}
\ee
and  $k_y$ is  the reciprocal lattice vector in $y$-direction. 
We find the uniformly extended states at $|k_y| = \frac{2\pi}{3}$ as well as the strictly localized states 
at $|k_y| = \pi$.  In addition, the origin of the states with macroscopic 
degeneracy at $|k_y | =\pi $ with  $E=\pm 1 $ 
are explained.

The localization length of Armchair diverges and there is no edge states if it is  isotropic.
 We find edge states, however, if the system is anisotropic, i.e., three hoppings
$t_a$, $t_b$, and $t_c$ are not equal\cite{hasegawaKonno,hasegawaKohmoto}.

Some of our results have been reported previously 
in the tight-binding 
model\cite{klein,fujita,nakada,wakabayashi1999,kusa,hatsugai,ezawa2006,peres2006},
the effective Dirac equation\cite{brey2006,sasaki2006,abanin2006}
and the first principle calculations\cite{miyamoto1999,okada2001,okada2003,lee2005,son2006,son2006b}
For the isotropic tight-binding model, 
Klein\cite{klein} has obtained  the condition for the existence of the edge states for the 
bearded edge.
Fujita et al.\cite{fujita} obtained the conditions for the zigzag and armchair edges.
However, there has been no studies of the anisotropic cases.

Due to the Dirac zero modes Jahn-Teller effect could take place in this system.
Dynamical breaking of the lattice symmetry would give rise to anisotropy.

\section{Tight-binding model}
Zigzag, Bearded, Zigzag-Bearded, and Armchair are shown  in  Figs. 2 (a), (b), (c), and (d).  
We pair a site on sublattice A and a site on sublattice B and denote the wave functions of 
a pair as $\psi_{n,m}$ and $\phi_{n, m} $ as shown in Fig. \ref{honey00},
where $n$ and $m$ are both integers or both half-integers. 
Then nearest neighbor hoppings 
give a tight-binding model
\begin{align}
 -t_a \psi_{n,m} - t_b \psi_{n+\frac{1}{2},m-\frac{1}{2}} - t_c \psi_{n+\frac{1}{2},m+\frac{1}{2}} &
 = E \phi_{n,m} 
\nonumber \\
 -t_a \phi_{n,m} - t_b \phi_{n-\frac{1}{2},m+\frac{1}{2}} - t_c \phi_{n-\frac{1}{2},m-\frac{1}{2}} 
 &= E \psi_{n,m} ,
\label{tbm0}
\end{align}
where $t_a$  is an intra-pair hoppings between $\psi_{n.m}$ and $\phi_{n,m}$,
$t_b$ and $t_c$ and  are  inter-pair hoppings.

 \section{Zigzag and Bearded }
 The zigzag edge and bearded edge can appear in the left edge or right edge. 
If the left edge is formed by the sublattice A or B, it is  bearded or zigzag, respectively.
If the right edge  is formed by the sublattice A or B, it is  zigzag or bearded, respectively.

We have edges in $y$-direction as shown in Figs. 2(a)-(c) and impose 
the periodic boundary condition,  $\psi_{n,m+L_y} = \psi_{n,m}$ and 
$\phi_{n,m+L_y} = \phi_{n,m}$ in $y$-direction, where $L_y$ is an 
 integer. Then one can write 
\be 
\psi_{n,m} = \exp(ik_ym) \psi_n \nonumber \\
\phi_{n, m} = \exp(ik_ym)\phi_n,
\ee
 where $k_y = \frac{2\pi j}{L_y}$ and  $j= 1, \ldots , L_y$ and (\ref{tbm0}) is written
\be
 \psi_n + t_1 \psi_{n+\frac{1}{2}} &=&- \frac{E}{t_a} \phi_n
\nonumber \\
 \phi_n + t_2 \phi_{n-\frac{1}{2}} &=&-\frac{E}{t_a} \psi_n,
  \label{tb}
\ee
where
\be
t_1= \frac{ t_b e^{-i\frac{k_y}{2}}+t_c e^{i\frac{k_y}{2}}}{t_a}
\label{tky1} \\
t_2=t_1^*= \frac{t_b e^{i\frac{k_y}{2}}+t_c e^{-i\frac{k_y}{2}}}{t_a}.
\label{tky2}
\ee
We define
\be
t = |t_1| = |t_2|=\frac{\sqrt{t_b^2+t_c^2+2t_bt_c\cos k_y}}{t_a}.
\label{tt}
\ee
When $t_b = t_c$, we have
\be
t_1 = \frac{2t_b}{t_a} \cos \frac{k_y}{2}.
\ee

\subsection {Macroscopically degenerate states and the strictly localized states}
If $t_b = t_c$ and $ |k_y| = \pi$, $t_1$ and $t_2$ in (\ref{tky1}) and (\ref{tky2}) vanish and (\ref{tb}) becomes 
\be
\psi_{n} &=& -\frac{E}{t_a} \phi_n
\nonumber 
\\
\phi_{n} &=& -\frac{E}{t_a} \psi_n.
\label{eqpair}
\ee
There is no inter-pair coupling and each pair is decoupled from others. 
This leads to  macroscopic degeneracy. These are bulk  states
$E=\pm t_a$
for Zigzag, Bearded, 
and Zigzag-Bearded as shown in 
Figs 3, 7, and 8, 
respectively.

If we have $E=0$ in addition,  pairs ($\psi_{n}$, $\phi_{n}$) vanish as seen from (\ref{eqpair}).  
Only non-vanishing wave functions are the unpaired ones at the edges. 
For Zigzag, the unpaired wave functions are $\phi_{1/2}$ and $\psi_{L_x+1/2}$.  See Fig. 2(a). 
Thus we have strictly localized states $\phi_{1/2}$ at the left edge and $\psi_{L_x+1/2}$ at the right edge. 
(These states are extended in the $y$-direction.) 
For Bearded, there is no unpaired state as shown 
in Fig. 2(b), and there is no strictly localized state.    
 For Zigzag-Bearded, $\phi_1$ is unpaired as shown in Fig. 2(c).
This is the strictly localized state at the left edge.
(This state is extended in the $y$-direction.)

\subsection {Edge States}
In the followings we implicitly assume the appropriate thermodynamic limit.
If $E=0$,  (\ref{tb})   is reduced to
\be
 \psi_n  &=&- t_1 \psi_{n+\frac{1}{2}} \nonumber \\
 \phi_n  &=&  - t_2 \phi_{n-\frac{1}{2}}.
 \label{tbe2}
\ee
There is no intra pair coupling, namely $\psi_n$'s on sublattice A and $\phi_n$'s on sublattice B are decoupled.

\subsubsection{\protect\ Zigzag}
As shown in Fig.2(a), the left edge has $\phi_{\frac{1}{2}}$ on sublattice B 
and the right edge has $\psi_{L_x+\frac{1}{2}}$ on sublattice A.
The boundary condition on the left edge is to 
add a fictitious sites with $\psi_{\frac{1}{2}}=0$. In the same manner, the boundary condition on the right edge is to add a fictitious sites with $\phi_{L_x+\frac{1}{2}}=0$.
Thus the boundary conditions for Zigzag are 
\begin{align}
 \psi_{\frac{1}{2}} &=0
\label{zigbc1}\\
 \phi_{L_x+\frac{1}{2}}& =0.
\label{zigbc2}
\end{align}
From  (\ref{tbe2})  we obtain
\be
\psi_{L_x-n}&=&(-t_1)^{2n+\frac{1}{2}}\psi_{L_x+\frac{1}{2}}
\label{edge1} \\
 \phi_{n}&=&  (- t_2)^{2n-1} \phi_{\frac{1}{2}}.
 \label{edge2}
 \ee
For the system with finite $L_x$, the solutions above are not exact.
If $t < 1$, however, (\ref{edge1}) and (\ref{edge2}) satisfy the boundary conditions in the limit
$L_x \rightarrow \infty$ and
these give  the right edge states on sublattice A and
the left edge states on sublattice B, respectively.
Even if the system size is finite, we have the edge states with exponentially small $E$, when the edge states on the sublattices A and B coexist with negligibly small mixing to satisfy the boundary conditions.

The localization lengths for both sublattices are the same  and given by
\be
\xi = \frac{1}{2 |\log{t}|}.
\label{length}
\ee
From (\ref{tt}), the condition for the existence of the edge states, $t<1$, is given by
\begin{equation}
\cos k_y <\frac{t_a^2 - t_b^2 -t_c^2}{2 t_b t_c}.
\label{cond-ky-zigzag}
\end{equation}
Thus we have edge states for all the values of $k_y$ if $  t_b + t_c < t_a$.
There are no edge states if $ | t_b - t_c | > t_a$.

\begin{figure}[t]
\begin{center}
\includegraphics[width=80mm]{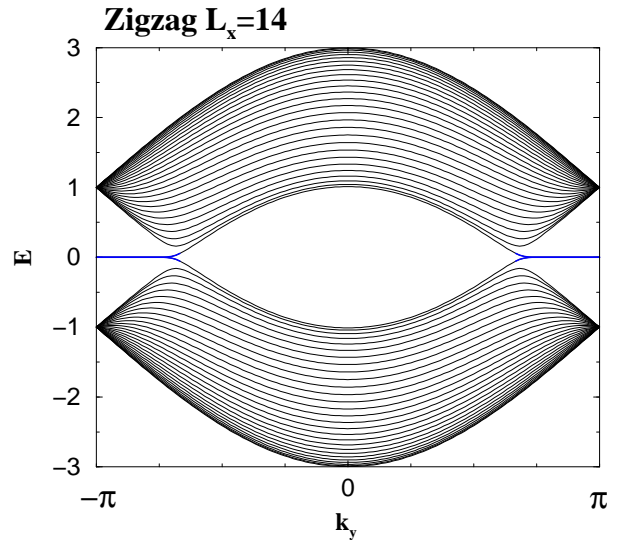}
\caption {Energy spectrum for Zigzag for $L_x=14$. The zero energy modes for 
the edge states are on the blue lines.}
\label{zig_spe}
\end{center}
\end{figure}

For $t_a=t_b=t_c$, edge states exist if
$|k_y|>\frac{2\pi}{3}$.
The zero energy modes for these edge states are seen in  Fig.~\ref{zig_spe}. 

\begin{figure}[t]
\begin{center}
\includegraphics[width=75mm]{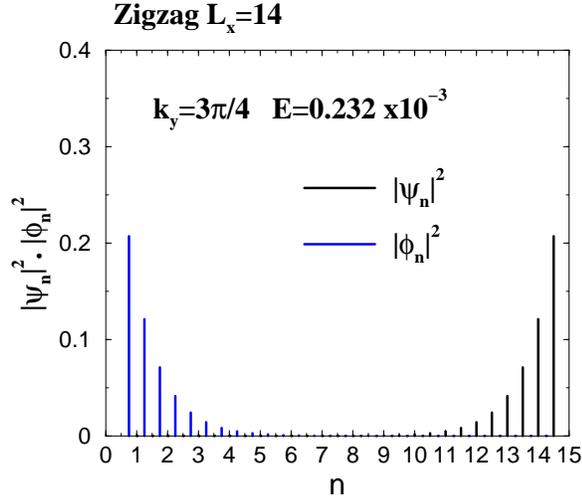}
\caption {An example of the edge states.}
\label{zigEVk34}
\end{center}
\end{figure}

An example of edge states are shown in Fig. \ref{zigEVk34} 

\subsubsection{Bearded}
\begin{figure}[t]
\begin{center}
\includegraphics[width=80mm]{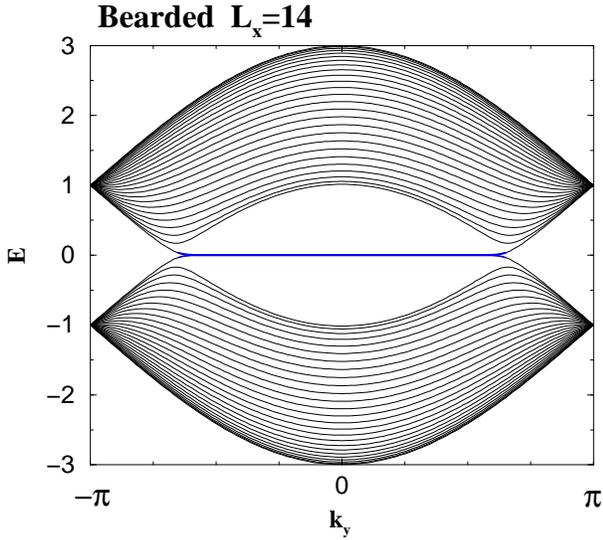}
\caption {Energy spectrum for Bearded. The zero energy mode for 
the edge states are on the blue line.} 
\label{be_spe}
\end{center}
\end{figure}

As shown in Fig. 2(b), the left edge has $\psi_1$ on sublattice A and 
the right edge has $\phi_{L_x}$ on sublattice B.
The boundary conditions are given by 
\begin{align}
\psi_{L_x +\frac{1}{2}} &= 0 \nonumber \\
\phi_{\frac{1}{2}} &=0 .
\end{align} 
In this case write (\ref{tbe2}) as
\be
  \psi_n &=&\left( -\frac{1}{t_1}\right)^{2n-2}\psi_1 \nonumber   \\
 \phi_{L_x-n} &=& \left( -\frac{1}{t_2}\right)^{2n}\phi_{L_x} .
\label{beard-edge2}
\ee
If $t>1$,
these are edge states with the  localization length 
\be
\xi = \frac{1}{2 \log{t}}.
\label{length2}
\ee
For $t_a=t_b=t_c$, edge sates exist if $|k_y|<\frac{2\pi}{3}$.
The energy spectrum in the isotropic case is plotted in Fig.~\ref{be_spe}. 

\subsubsection{Zigzag-Bearded}
As shown in Fig. \ref{fig2}(c), the left edge is zigzag with $\phi_{\frac{1}{2}}$  on sublattice B. 
The right edge is bearded with $\phi_{L_x}$ also on sublattice B.  
The boundary conditions are given by
\begin{align}
 \psi_{\frac{1}{2}} &= 0\\ 
 \psi_{L_x+\frac{1}{2}} &=0 .
\end{align} 
These boundary conditions give $\psi_n=0$. Thus there are no edge states on sublattice A.
There are no boundary condition for $\phi_n$.
From (\ref{tbe2}) we have
\be
\phi_n&=&(-t_2)^{2n-1}\phi_{\frac{1}{2}}.
\label{zb1} 
\ee
This is left edge states if $t<1$.
On the other hand  write
\be
 \phi_{L_x-n}=  \left(- \frac{1}{t_2}\right)^{2n} \phi_{L_x},
 \label{zb2}
 \ee
then this  gives right edge state  if $t>1$.

\begin{figure}[t]
\begin{center}
\includegraphics[width=80mm]{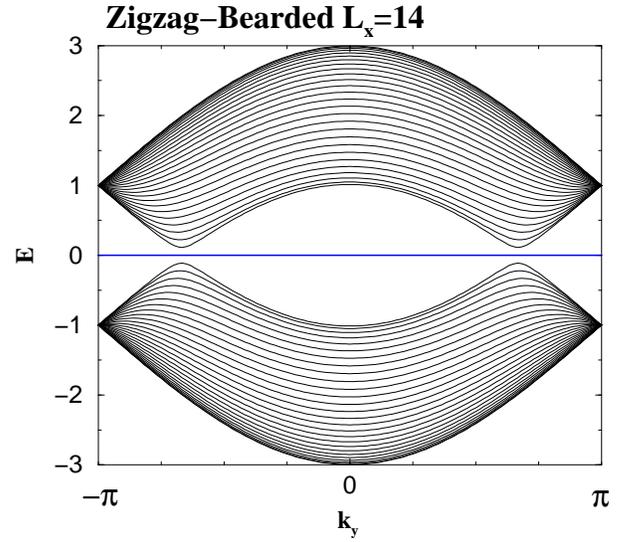}
\caption {Energy spectrum for Zigzag-Bearded for $L_x=40$. The zero energy mode for 
the edge states are on the blue line.}
\label{z_b_spe}
\end{center}
\end{figure}

The energy spectrum in the isotropic case, $t_a=t_b=t_c$, 
are shown in Fig.~\ref{z_b_spe}. 
The  edge states  are on the blue line which has the full length 
from $-\pi$ to $ \pi$.  

\section{Armchair}
We have  the periodic boundary condition in $x$-direction,
$\psi_{n + L_x, m} = \psi_{n,m}$ and $\phi_{n+ L_x, m} = \phi_{n,m}$, as shown in Fig. 2(d).
So write 
\be
\psi_{n,m} &=& \exp(ik_xn) \psi_m \nonumber \\
\phi_{n, m} &=& \exp(ik_x n)\phi_m,
\label{bcs_arm}
\ee 
where 
$k_x = \frac{2\pi j}{L_x}$ and  $j= 1, \ldots , L_x$. 
The boundary conditions at the edges are
\begin{align}
\psi_{\frac{1}{2}}&=\phi_{\frac{1}{2}}=0
\nonumber \\
\psi_{L_y+\frac{1}{2}}&=\phi_{L_y+\frac{1}{2}}=0. 
\label{bc1}
\end{align}

 Let us consider the case where $E= 0$ in which  $\psi$'s and $\phi$'s 
are decoupled and satisfy 
\be
-t_a \psi_m - e^{i\frac{k_x}{2}} (t_b\psi_{m-\frac{1}{2}} + t_c  \psi_{m+\frac{1}{2}}) & =& 0 \nonumber \\
-t_a \phi_{m} - e^{-i\frac{k_x}{2}}(t_b \phi_{m+\frac{1}{2}} + t_c \phi_{m-\frac{1}{2}} ) & =& 0.
\label{a_eq5b}
\ee
Put
\be
\psi_m=z^{2m},
\ee
then from (\ref{a_eq5b}) we have two solutions for $z$ which satisfy
\be
z_{\pm} &=& \frac{1}{2} \left[ -\frac{t_a}{t_c} e^{-ik_x/2} \pm \sqrt{ (\frac{t_a}{t_c})^2 e^{-ik_x} - 4\frac{t_b}{t_c}} \, \right]
\nonumber \\
z_+ z_- &=& \frac{t_b}{t_c} \nonumber \\
z_+ + z_- &=& -\frac{t_a}{t_c} e^{-ik_x/2}.
\label{zroot}
\ee
In terms of these, the solution for (\ref{a_eq5b}) with 
the boundary condition $\psi_{\frac{1}{2}} =0$ is given by
\begin{equation}
\psi_m=\frac{z_+^{2m-1}-z_-^{2m-1}}{z_{+}-z_{-}} \psi_1.
\label{armbottom}
\end{equation}
This satisfies the boundary condition in the limit $L_y \rightarrow \infty$ if
\be
|z_{+}| < 1 \; \mbox{and} \; |z_{-}| <1.
\label{zpm1}
\ee
They are edge states localized near the bottom. In order (\ref{zpm1}) be satisfied, 
\be
t_b < t_c
\ee
 is required as seen from (\ref{zroot}).

In a similar manner, edge states localized near the top are possible if
\be
|z_{+}| >1\; \mbox{and}\; |z_{-}| >1. 
\label{zpm2}
\ee
As seen from (\ref{zroot})
\be
 t_b > t_c
 \ee
is required.

For analysis of $\phi$'s, we only need to replace $k_x$ by $-k_x$ and $t_b$ and $t_c$. This symmetry can be seen in Fig. (1) and also in Eq. (\ref{a_eq5b}). Thus we obtain essentially the same conditions.

\begin{figure}[t]
\begin{center}
\includegraphics[width=80mm]{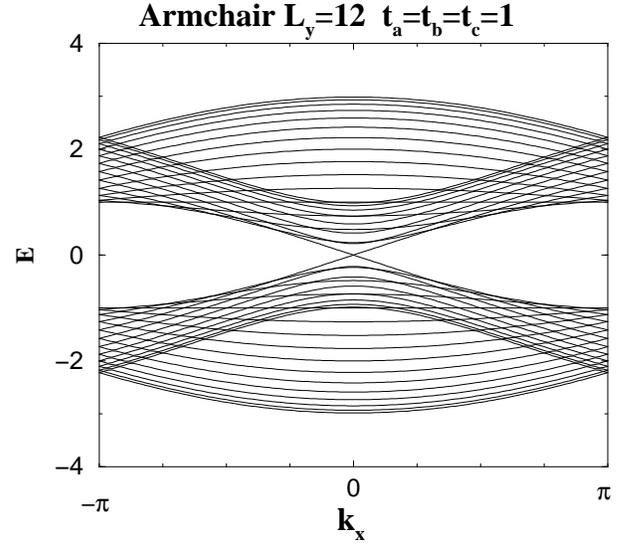}
\caption {Energy spectrum for Armchair in the isotropic case. Neither (\ref{zpm1}) nor (\ref{zpm2}) are satisfied in this case. There is no edge state.}
\label{a_spe2}
\end{center}
\end{figure}

\begin{figure}[h]
\includegraphics[width=80mm]{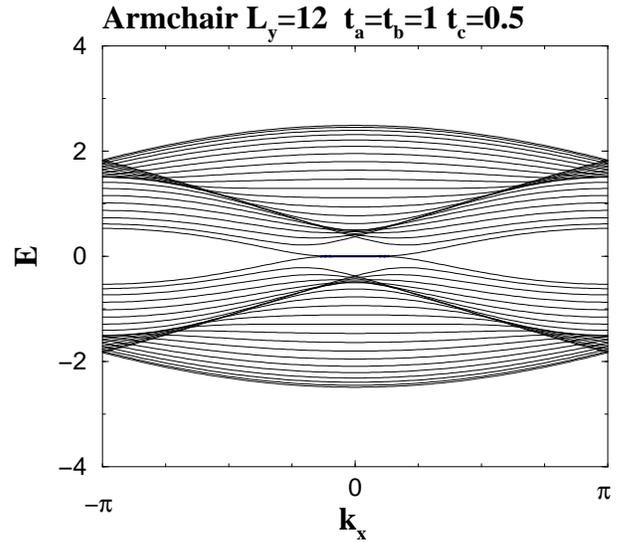}
\caption{Energy spectrum for Armchair in an anisotropic case. The condition (\ref{zpm2}) is satisfied for certain $k_x$'s  and edge states exist on the blue line. }
\label{fig4z}
\end{figure}

See  Figs. \ref{a_spe2} and \ref{fig4z} for examples of the  energy spectrum.

\clearpage

\section{acknowledgement}
We thank T. Aoyama  for help with a computer calculation.


\end{document}